\documentclass[prl,aps,twocolumn,superscriptaddress,
longbibliography]{revtex4-2}

\usepackage{tikz}
\usepackage{graphicx}
\usepackage{array}[=2016-10-06]
\usepackage{dcolumn}
\usepackage{bm}
\usepackage{enumitem}

\usepackage{yfonts}
\usepackage[bottom]{footmisc}
\usepackage[utf8]{inputenc}
\usepackage{amssymb, amsthm, tabto, epigraph, mathtools, dsfont, verbatim, slashed, mathrsfs}

\usepackage{graphicx, wrapfig}
\usepackage{enumitem}
\usepackage{relsize}
\usepackage[compatibility=false]{caption}
\usepackage{subcaption}
\usepackage{csquotes}
\usepackage{hyperref}
\usepackage{physics}
\usepackage{tabularx}
\usepackage[dvipsnames]{xcolor}
\usepackage{color}
\usepackage{array}
\usepackage{amsmath,amssymb}
\usepackage{array,booktabs,tabularx}
\usepackage[T1]{fontenc}
\usepackage[utf8]{inputenc}
\usepackage{lmodern}
\usepackage{microtype}
\usepackage{amsmath,amssymb,mathtools}
\usepackage{booktabs}
\usepackage{enumitem}
\usepackage[dvipsnames]{xcolor}
\usepackage{fancyhdr}
\usepackage{hyperref}
\usepackage[normalem]{ulem}
\usepackage[T1]{fontenc} 
\usepackage{blkarray}
\usepackage{relsize}

\usepackage{filecontents}

\usepackage{yfonts}
\usepackage[bottom]{footmisc}
\usepackage[utf8]{inputenc}
\usepackage{amssymb, amsthm, tabto, epigraph, mathtools, dsfont, verbatim, slashed, mathrsfs}
\usepackage{graphicx, wrapfig}
\usepackage{enumitem}
\usepackage{relsize}
\usepackage{subcaption}
\usepackage{csquotes}
\usepackage{hyperref}
\usepackage{physics}

\usepackage[dvipsnames]{xcolor}
\usepackage{color}

\usepackage[T1]{fontenc}
\usepackage[utf8]{inputenc}
\usepackage{lmodern}
\usepackage{microtype}
\usepackage{amsmath,amssymb,mathtools}
\usepackage{booktabs,array}
\usepackage{enumitem}
\usepackage[dvipsnames]{xcolor}
\usepackage{fancyhdr}
\usepackage{hyperref}

\newcolumntype{Y}{>{\raggedright\arraybackslash}X}

\newcommand\IZ{\mathbb{Z}}

\newcommand\CA{\mathcal{A}}
\newcommand\CB{\mathcal{B}}

\newcommand\CD{\mathcal{D}}

\newcommand\CL{\mathcal{L}}

\newcommand\CP{\mathcal{P}}

\newcommand\Spin{\operatorname{Spin}}

\newcommand{\ii}{\mathsf{i}}

\usepackage{amsmath}
\usepackage{empheq}
 
\setlist{nolistsep}
\hypersetup{
    colorlinks=true,
    linkcolor=Mahogany,
    filecolor=magenta,      
    urlcolor=blue,
}

\newlength\dlf  

\allowdisplaybreaks

\setlist{nolistsep}
\hypersetup{
    colorlinks=true,
    linkcolor=Maroon,
    filecolor=Maroon,      
    urlcolor=Maroon,
    citecolor=Maroon
}

\theoremstyle{remark}

\theoremstyle{definition}

\usepackage{amssymb}
\usepackage{tikz}
\usetikzlibrary{shapes,arrows,cd,chains,decorations.markings,decorations.pathmorphing,calc}
\tikzset{
->-/.style args={#1rotate#2}{decoration={markings, mark=at position #1 with {\arrow[scale=1.5,rotate = #2 ]{stealth}}}, postaction={decorate}}
}
\tikzset{curve/.style={settings={#1},to path={(\tikztostart)
    .. controls ($(\tikztostart)!\pv{pos}!(\tikztotarget)!\pv{height}!270:(\tikztotarget)$)
    and ($(\tikztostart)!1-\pv{pos}!(\tikztotarget)!\pv{height}!270:(\tikztotarget)$)
    .. (\tikztotarget)\tikztonodes}},
    settings/.code={\tikzset{quiver/.cd,#1}
        \def\pv##1{\pgfkeysvalueof{/tikz/quiver/##1}}},
    quiver/.cd,pos/.initial=0.35,height/.initial=0}
\tikzset{tail reversed/.code={\pgfsetarrowsstart{tikzcd to}}}
\tikzset{2tail/.code={\pgfsetarrowsstart{Implies[reversed]}}}
\tikzset{2tail reversed/.code={\pgfsetarrowsstart{Implies}}}

\newmuskip\pFqskip
\mathchardef\pFcomma=\mathcode`, 

\graphicspath{{./images/}}

\usepackage{braket}

\usepackage[export]{adjustbox}

\usepackage{caption}
\newcommand{\SMref}[1]{\hyperref[#1]{SM~\ref*{#1}}}

\usetikzlibrary{calc}  

\newcommand{\rC}{\mathrm{C}}
\newcommand{\rB}{\mathrm{B}}

\usetikzlibrary{arrows.meta}

\newcommand{\tG}{\widetilde{G}}

\begin{document}

\preprint{APS/123-QED}

\title{Anyon Proliferation and Anyon Superconductivity in Higgsing Transitions via \\ Conformal Embeddings}

\author{Diego Garc\'ia-Sep\'ulveda,}
\email{dgarciasepulveda@fas.harvard.edu}
\affiliation{Society of Fellows, Harvard University, Cambridge, MA 02138, USA}
\affiliation{Department of Physics, Harvard University, Cambridge, MA 02138, USA}

\author{Da-Chuan Lu}
\email{dachuan_lu@fas.harvard.edu}
\affiliation{Department of Physics, Harvard University, Cambridge, MA 02138, USA}
\affiliation{Department of Physics and Center for Theory of Quantum Matter, University of Colorado, Boulder, CO 80309, USA}

\begin{abstract}

We discuss (2+1)d Chern-Simons Higgsing transitions based on conformal embeddings and their relation to the proliferation of anyons. These transitions preserve or enlarge the intrinsic topological order despite reducing the gauge group. When the topological order is enlarged, the two phases differ by anyon condensation, and the transition takes an Abelian topological order to a non-Abelian one. In the $SO(N)_{2} \hookrightarrow SU(N)_{1}$ family with $N \geq 3$, the two semiclassical regimes have equivalent intrinsic topological orders but distinct realizations of a global $U(1)$ symmetry. For odd $N$, these are non-Abelian realizations of transitions from fermionic $1/N$ Laughlin states to charge-$2e$ anyon superconductivity coexisting with the same chiral $\mathbb{Z}_{N}$ topological order. The scalar representations are always non-integrable with respect to the UV Chern-Simons level and therefore do not directly label anyons of the un-Higgsed phase, despite defining Wilson lines in the UV. Using heuristic adjoint screening, we propose candidate anyons in the un-Higgsed phase, while branching under the Higgs subgroup reveals channels corresponding to condensable anyons of the Higgs phase. We further construct a $\mathrm{Spin}(16)_{1} \to \mathrm{Spin}(9)_{2}$ transition admitting a condensable algebra that contains a non-Abelian anyon but not all of the allowed gaugable channels in its self-fusion.
\end{abstract}

\maketitle


\textit{\textbf{Introduction.} ---} Topologically ordered phases support anyons with fusion and braiding properties beyond those of ordinary bosons and fermions. Understanding how these excitations drive phase transitions is a central question in quantum many-body physics. In particular, when anyons become light and proliferate, their interactions can produce new topological orders. Even ordinary bosons illustrate this possibility: in a magnetic field, depending on their interactions and filling, they can form a conventional condensate or instead realize distinct fractional quantum Hall states. What determines the outcome of anyon proliferation, and how is that outcome related to the anyons driving the transition?

These questions require distinguishing algebraic anyon condensation from dynamical anyon proliferation. A bosonic topological order is described by a modular tensor category (MTC), within which condensable algebras characterize the possible outcomes of algebraic anyon condensation \cite{Bais:2008ni,Kong:2013aya}. This algebraic description does not, by itself, determine which phase a microscopic system favors. More generally, anyon proliferation need not implement a condensable algebra in the MTC of the un-Higgsed phase: its outcome depends on the interactions and can include emergent topological order, spontaneous symmetry breaking, or other collective behavior. Recent approaches to proliferation transitions, including ``stack-and-condense'' and symmetry topological field theory (SymTFT) constructions, have sharpened this distinction and clarified the respective roles of algebraic constraints and dynamical input in determining the resulting phase~\cite{zhang2024hierarchy,Cordova:2025eim,Cheng:2026qax,Schafer-Nameki:2026hgx,Cheng:2026hgu}. Related studies have investigated competition between proliferating anyons in systems with self-duality symmetry~\cite{Shi:2024pem,Ji:2026yfj,Dumitrescu:2026vre,Lu:2026fid}.

Chern--Simons (CS) matter theories provide concrete settings in which to investigate anyon proliferation. Indeed, a gauge-charged scalar defines an endpoint for Wilson lines, and as its mass is tuned the corresponding excitation becomes light. With an appropriate potential, this drives a Higgsing transition that generically separates phases with distinct topological order.

Particularly interesting Higgsing transitions are those in which the embedding of gauge groups is a \textit{conformal embedding}. The two phases are then generically related by anyon condensation, making such transitions a straightforward setting in which to understand the interplay between dynamical anyon proliferation and the algebraic procedure of anyon condensation. Conformal embeddings have also been studied recently in a variety of contexts, including (2+1)d Higgsing and proliferation transitions~\cite{Komargodski:2020mxz, Yu:2021zmu, Delmastro:2021otj, Diatlyk:2023fwf, Cordova:2023jip, Cordova:2024goh, Cordova:2024nux, Lim:2024qis, Moore:2026lmq, Schafer-Nameki:2026hgx, Cheng:2026hgu}.

Of particular interest are conformal embeddings whose condensation involves only Abelian bosons, which we call \textit{Abelian conformal embeddings} (ACEs). Each implies a duality of CS theories identifying the topological orders before and after Higgsing, so when the Higgs stabilizer realizes precisely this quotient, the two regimes actually support equivalent topological orders.

To understand these transitions between phases with equivalent topological order, we introduce a global $U(1)_{B}$ symmetry and distinguish the phases by its realization. When $U(1)_{B}$ represents electric-charge conservation, this construction in fact realizes a transition of the kind studied in the seemingly separate context of anyon superconductivity~\cite{Laughlin:1988ft,Chen:1989xs,lee1989anyon}. This subject has recently attracted renewed interest~\cite{
Shi:2024cgb,ShiSenthil2025,Wang:2026ziw,
Shi:2026oxz,Mehta:2026vnh,SeoLeeCho2026}, motivated in part by experimental signatures of superconductivity near fractional quantum anomalous Hall states in twisted bilayer $\mathrm{MoTe}_{2}$~\cite{xu2025signatures}.

We illustrate our observations with a simple family of orthogonal-unitary embeddings: $SO(N) \subset SU(N)$ \cite{Schellekens:1986mb,Bais:1986zs,Aharony:2016jvv,Lim:2024qis,Zhang:2025xkw}. More precisely, we couple $SU(N)_{1}$ to a scalar in the rank-two symmetric representation. An isotropic configuration of the scalar Higgses the gauge group to $SO(N)_{2}$, which is, however, isomorphic to $SU(N)_{1}$ itself: $SO(N)_{2} \cong SU(N)_{1}$. A gauge-invariant $U(1)_B$-charged composite nevertheless condenses and spontaneously breaks $U(1)_{B}$, distinguishing the Higgs phase from the un-Higgsed phase. With the appropriate fermionic and invertible sectors included, the odd-$N$ examples describe transitions from Laughlin states to charge-$2e$ superconductors coexisting with chiral $\mathbb{Z}_{N}$ topological order, namely SC$^{*}$ phases~\cite{senthil2000z,tang2013scs}. The present construction thus provides a manifestly non-Abelian realization of anyon superconductivity transitions. For even $N$, interpreting $U(1)_{B}$ as spin rotations instead yields magnetic ordering coexisting with chiral spin liquid.

We further study the Higgsing transition from $\Spin(16)_1$ to $\Spin(9)_2$, in which non-Abelian topological sectors emerge from an Abelian topological order of the un-Higgsed phase. The transition from the Higgsed phase to the un-Higgsed phase illustrates that the phases are related by non-Abelian anyon condensation, with a condensable algebra that contains a non-Abelian anyon without including every gaugable channel in its self-fusion~\cite{BaisSchroersSlingerland2003,Kong:2013aya,LanWangWen2015,Lu:2026fid}.

Additional examples where the phases differ by anyon condensation appear in the End Matter, where we also discuss how the scalar content in $\Spin(8)_1$ determines whether an additional toric-code sector accompanies the three-fermion order. In the Supplemental Material, exceptional-group constructions yield either equivalent Fibonacci order on both sides or invertible phases with the same $E_8$ Kitaev state and no nontrivial anyons on either side.

\textit{\textbf{Setup.} ---} We consider CS-matter theories in $(2+1)$ spacetime dimensions. The gauge group is denoted by $G$ and the level by $k$, and we always take $G$ to be simply connected. The CS term defined by this pair is denoted by $G_{k}$. We also introduce a scalar field $\phi_{\mathbf{R}}$ transforming in a representation $\mathbf{R}$ of $G$. Unless otherwise noted, $\mathbf{R}$ is irreducible. When $\mathbf{R}$ is complex, $\phi_{\mathbf{R}}$ is a complex scalar.

In addition to the minimal coupling between the scalar and the CS gauge fields, we introduce a scalar potential $V(\phi_{\mathbf{R}})$ with couplings such that the Higgs condensate has the desired stabilizer subgroup. This potential is otherwise generic, and it includes in particular a mass term:
\begin{equation}
    V(\phi_{\mathbf{R}}) = m^{2}|\phi_{\mathbf{R}}|^{2} + \cdots,
\end{equation}
where the ellipsis denotes higher order interactions. The mass term is responsible for the Higgs transition between the two semiclassical regimes at large $|m^{2}|$  \footnote{Our analysis determines the semiclassical phases and candidate proliferation channels. It does not establish the order of the transition or exclude intermediate phases.}. For large $m^{2} > 0$ the scalar field has vanishing expectation value and can be trivially integrated out. At long distances the system is described by the CS Lagrangian $G_{k}$. For large $m^{2} < 0$ the scalar field condenses and one finds a Higgs phase described by a smaller gauge group: the stabilizer $F \subset G$ of the scalar VEV, $f \phi_{0} = \phi_{0}$ for $f \in F$. It will be important in what follows to recall that the stabilizer subgroup need not be simply connected. The level of the infrared CS theory is an integer multiple of the UV level $k$: $\tilde{k} = \ell k$, where $\ell$ is the embedding index of $F$ in $G$ \footnote{The index of embedding can be computed by decomposing any representation $\mathbf{R}_{G}$ of $G$ in terms of representations $\mathbf{R}_{F}$ of the subgroup $F$ and taking the quotient of Dynkin indices: $\ell = I(\mathbf{R}_{F})/I(\mathbf{R}_{G})$.}. 

The generic Higgsing RG flow looks as follows:
\begin{equation}
\begin{tikzpicture}[
  x=1pt,y=1pt,
  every node/.style={
    inner sep=0pt,
    outer sep=0pt,
    anchor=base
  },
  phase/.style={font=\fontsize{10}{12}\selectfont},
  note/.style={font=\fontsize{9}{11}\selectfont},
  branch/.style={
    -{Stealth[length=3.4pt,width=2.5pt]},
    line width=0.4pt
  }
]

  \node[phase] at (0,0) {$G_k+\phi_{\mathbf R}$};

  \draw[branch] (-14,-8) -- (-48,-27);
  \draw[branch] (14,-8) -- (48,-27);
  \node[note] at (-53,-15) {$m^2<0$};
  \node[note] at (53,-15) {$m^2>0$};

  \node[phase] at (-48,-37) {Higgs};
  \node[phase] at (-48,-47) {$F_{\tilde{k}}$};
  \node[phase] at (48,-37) {Un-Higgsed};
  \node[phase] at (48,-47) {$G_k$};
\end{tikzpicture}
\label{higgsing_flow_Fk}
\end{equation}
The TQFTs $F_{\tilde{k}}$ and $G_{k}$ are generally distinct: they typically differ in their central charge or their anyon content. The Higgsing then models the transition between these two topologically ordered phases.

\textit{Conformal Embeddings and Duality.} We will be interested in group embeddings whose associated embedding of affine Lie algebras is \textit{conformal}. More precisely, an embedding of affine Lie algebras $\mathfrak{h}_{\tilde{k}} \hookrightarrow \mathfrak{g}_{k}$ is \textit{conformal} if the central charges match: $c_{\mathfrak{h}_{\tilde{k}}} = c_{\mathfrak{g}_{k}}$ \footnote{In the context of 2D CFT, an embedding of affine Lie algebras $\mathfrak{h}_{\tilde{k}} \hookrightarrow \mathfrak{g}_{k}$ is called conformal when the Sugawara stress tensors constructed from the currents satisfy $T_{\mathfrak{h}_{\tilde{k}}} = T_{\mathfrak{g}_{k}}$. Since we are interested here in TQFTs, the condition on the central charges suffices for our purposes.}. It is straightforward to prove that this can hold only if $k = 1$~\cite{DiFrancesco:1997nk}, which we assume henceforth.

For a conformal embedding, the two CS theories based on these affine Lie algebras are related by non-Abelian anyon condensation~\cite{Bais:2008ni, davydov2013witt, Kong:2013aya, Yu:2021zmu, Cordova:2023jip}:
\begin{equation} \label{ConformalEmbeddingDualityGlobalForm}
    H_{\ell}/\mathcal{A} \cong G_{1},
\end{equation}
where $H$ is a Lie group with algebra $\mathfrak{h}$ \footnote{Eqn. \eqref{ConformalEmbeddingDualityGlobalForm} is a manifestation of the fact that the categories associated with a conformal embedding are Witt equivalent \cite{davydov2013witt}.}. We emphasize that $H$ is not necessarily simply connected. That is, we track the actual embedding of groups rather than only that of the algebras \footnote{Denoting by $H^{\mathrm{sc}}$ the simply connected Lie group with Lie algebra $\mathfrak{h}$, $H_{\ell}$ may itself be obtained as a condensation of $H^{\mathrm{sc}}_{\ell}$ by some subgroup $Z(H^{\mathrm{sc}})'$ of the center $Z(H^{\mathrm{sc}})$: $H_{\ell} \equiv H^{\mathrm{sc}}_{\ell} / Z(H^{\mathrm{sc}})'$.}.

In \cite{Cordova:2023jip}, the relation \eqref{ConformalEmbeddingDualityGlobalForm} was interpreted as a general statement about dualities of TQFTs. For some conformal embeddings, $\mathcal{A}$ turns out to be trivial, in which case Eqn. \eqref{ConformalEmbeddingDualityGlobalForm} reduces to the usual notion of duality of CS theories:
\begin{equation} \label{AbelianDuality}
    H_{\ell} \cong G_{1}.
\end{equation}
Embeddings of this class are precisely what we referred to as ACEs in the Introduction \footnote{We emphasize that the duality statements \eqref{ConformalEmbeddingDualityGlobalForm}--\eqref{AbelianDuality} are kinematical statements about (2+1)d TQFTs. That is, they are a priori unrelated to any phase transition or other dynamical process that produces non-trivial topological order at long distances.}. Note the resemblance to the more standard level-rank dualities \cite{Mlawer:1990uv, Naculich:1990pa, Naculich:2007nc, Hsin:2016blu, Aharony:2016jvv, Cordova:2017vab, Cordova:2018qvg, Cordova:2023jip}.

\textit{Consequences for Higgsing Transitions.} Conformal embeddings allow us to construct Higgsing transitions in which the two semiclassical phases are related by anyon condensation or, more surprisingly, have equivalent topological order. The key point is to keep track of the global forms of the gauge groups, since these determine the precise anyon content of the system at long distances. The discussion above involves two Lie groups whose global forms must be distinguished a priori:\\
\begin{itemize}
    \item The stabilizer group $F$ that determines the IR CS term after Higgsing in \eqref{higgsing_flow_Fk}. \\

    \item The Lie group $H$ in the dualities \eqref{ConformalEmbeddingDualityGlobalForm}-\eqref{AbelianDuality}. \\
\end{itemize}
These two groups actually coincide: for a generic potential and a suitable scalar representation, $F = H$ \footnote{This assumes a potential that Higgses $G$ to the connected group $F$ determined by the embedding, with no additional discrete factors from tunings of the potential or group-theoretical accidents. See Appendix A of \cite{Cordova:2018qvg} to see how to find such a potential.}. Eqn. \eqref{ConformalEmbeddingDualityGlobalForm} then ensures that the two topological orders are related by anyon condensation. For ACEs, \eqref{AbelianDuality} further ensures that the anyon content agrees across the two phases, so the Higgsing does not alter the topological order. In the latter case, we therefore obtain the phase diagram:
\begin{equation}
\begin{tikzpicture}[
    baseline=(current bounding box.center),
  x=1pt,y=1pt,
  every node/.style={
    inner sep=0pt,
    outer sep=0pt,
    anchor=base
  },
  phase/.style={font=\fontsize{10}{12}\selectfont},
  note/.style={font=\fontsize{9}{11}\selectfont},
  branch/.style={
    -{Stealth[length=3.4pt,width=2.5pt]},
    line width=0.4pt
  }
]
  
  \node[phase] at (0,0) {$G_1+\phi_{\mathbf R}$};

  \draw[branch] (-14,-8) -- (-48,-27);
  \draw[branch] (14,-8) -- (48,-27);
  \node[note] at (-53,-15) {$m^2<0$};
  \node[note] at (53,-15) {$m^2>0$};

  \node[phase] at (-48,-37) {Higgs};
  \node[phase] at (-48,-47) {$H_{\ell} \cong G_{1}$};
  \node[phase] at (0,-47) {};
  \node[phase] at (48,-37) {Un-Higgsed};
  \node[phase] at (48,-47) {$G_1$};
\end{tikzpicture}
\label{abelianconformalembeddingflow}
\end{equation}

\textit{\textbf{Zero-form Enrichment.} ---}
Global symmetries provide physical handles for probing and controlling topological orders. In fractional quantum Hall systems, anyons can carry fractional charge under the $U(1)_B$ symmetry associated with electric-charge conservation. We incorporate this enrichment by coupling the CS-matter theory to a background gauge field $B$. As we demonstrate below, spontaneous breaking of $U(1)_B$ distinguishes the two semiclassical regimes, realizing anyon superconductivity while preserving the intrinsic topological order.

For a fixed intrinsic topological order, $U(1)$ enrichment is specified by the fractional charges of anyons and the Hall response~\cite{ChengJian2022,Cheng:2022nji,seo2026anyonsc,Alam:2026dmh}. Since $U(1)$ is connected, it cannot permute anyon types. Phases with equivalent intrinsic topological order can be physically distinct through their symmetry realizations. Charge fractionalization is encoded by an Abelian anyon, the vison, associated with adiabatic insertion of $2\pi$ electromagnetic flux: its braiding with an anyon reproduces the corresponding Aharonov--Bohm phase~\cite{Barkeshli:2014cna,ChengJian2022,Cheng:2022nji}. In the Abelian $K$-matrix description, the charge vector labels this vison~\cite{WenZee1992,ChengJian2022}.

For a simply connected gauge group $G$ with center $\IZ_n$, the center one-form symmetry of $G_k$ is generated by Abelian lines that provide natural choices for the vison. A field-theoretic realization uses the extended gauge group
\begin{equation}
    \tG=\frac{G\times U(1)}{\IZ_n},
\end{equation}
where the quotient identifies the center of $G$ with the $\IZ_n$ subgroup of the added $U(1)$. We combine the dynamical $G$ gauge field and the electromagnetic background $B$ into a $\tG$ connection $\widetilde a$, whose induced connection on $\tG/G\simeq U(1)_B$ is fixed to $B$. For $G=SU(n)$, this amounts to using a $U(n)$ connection satisfying $\tr(\widetilde a)=B$, while its traceless part remains dynamical~\cite{Hsin:2016blu}. The quotient ties fractional electric charge to center charge: inserting $2\pi$ electromagnetic flux realizes the corresponding center simple current as the vison. We implement the background coupling through
\begin{equation}
    \mathcal L[B]
    =\mathcal L_{\mathrm{CS}}[\widetilde a]
    +\frac{1}{2\pi}c\,d\bigl(\tr(\widetilde a)-B\bigr)
    +\frac{p}{4\pi}B\,dB,
\end{equation}
where the compact $U(1)$ gauge field $c$ fixes $\tr(\widetilde a)$ to $B$, and the allowed background counterterm
shifts the Hall response by $p$. Matter representations are labeled by $(R,q)$, where $q$ is the integer charge under the added $U(1)$ and satisfies $q\equiv r_R\pmod n$. The center charge $r_R$ is the $n$-ality of $R$ for $G=SU(n)$.

\textit{\textbf{Unitary-Orthogonal Family.} ---}
The fermionic $1/3$ Laughlin state admits an $SU(3)_1$ description supplemented by a transparent-fermion sector and appropriate counterterms~\cite{Hsin:2016blu}\footnote{The bosonic $SU(3)_1$ can describe the Halperin (221) state.}. We focus here on the $SU(3)_1$ sector, leaving the full fermionic formulation to \SMref{SM:LevelRankDuality}. Using the $U(3)$ formulation introduced above, we add a complex scalar $\phi_R$ in the representation $(\mathbf6,2)$, where $\mathbf6=\mathrm{Sym}^2(\mathbf3)$ is its $SU(3)$ representation and $2$ is its charge under the central $U(1)$. The scalar is a symmetric $3\times3$ matrix transforming as $\phi_R\mapsto U\phi_R U^T$, with $U\in U(3)$. The resulting CS-Higgs Lagrangian is
\begin{equation}
\begin{aligned}
    \mathcal L[B]
    &=\tr[(D_\mu\phi_{\mathbf R})^\dagger D^\mu\phi_{\mathbf R}]-V(\phi_{\mathbf R})\\
    &\quad+\frac{1}{4\pi}
    \tr\left(\widetilde a\,d\widetilde a
    -\frac{2i}{3}\widetilde a^3\right)
    +\frac{1}{2\pi}c\,d(\tr\widetilde a-B),
\end{aligned}
\label{eq:su3_symmetric_csh}
\end{equation}
where
$D_\mu\phi_R=\partial_\mu\phi_R-i\widetilde a_\mu\phi_R-i\phi_R\widetilde a_\mu^T$.
The constraint $\tr\widetilde a=B$ gives the local decomposition $\widetilde a=a+\frac13B\mathbf1_3$, with $a$ traceless. 

The BF coupling gives the conserved current $\star j_B=-dc/(2\pi)$. The corresponding $U(1)_B$ symmetry operator on a closed surface $M_2$ is
\begin{equation}
    U_\alpha(M_2)
    =\exp\!\left(-\frac{\ii\alpha}{2\pi}\int_{M_2}dc\right).
    \label{eq:symmetry-surface}
\end{equation}
The fractional $U(1)_B$ charge for the fundamental line can be read off from the local decomposition above,
\begin{equation}
    W_{\mathbf3}(\gamma)
    =\tr\CP e^{\ii\oint_\gamma\widetilde a}
    =e^{\frac{\ii}{3}\oint_\gamma B}
    \tr\CP e^{\ii\oint_\gamma a},
\end{equation}
which identifies its $U(1)_B$ charge as $1/3$. The scalar $\phi_{(\mathbf6,2)}$ therefore carries physical charge $2/3$. We take $V(\phi_R)$ to preserve $U(1)_B$ and favor equal singular values in the ordered regime.

For large positive $m^2$, the massive scalar yields the intrinsic topological order $SU(3)_1$ with $U(1)_B$ symmetry. For large negative $m^2$, a representative minimum is $\phi_R=v\mathbf1_3$, whose stabilizer inside $SU(3)$ is precisely $SO(3)$. Restricting the Chern--Simons action gives orthogonal level two~\cite{Cordova:2025eim}, with $SO(3)_2 = \Spin(3)_2/\mathbb Z_2 \cong SU(2)_4/\mathbb Z_2 = SU(3)_1$.
The $\IZ_2$ quotient condenses the $j=2$ boson: $j=0$ and $j=2$ are identified with the vacuum, $j=1$ splits into two Abelian lines, and the half-integer-spin lines are confined~\cite{Aharony:2016jvv,Moore:1989yh,Hsin:2018vcg}.

Strikingly, the $m^2>0$ and $m^2<0$ regimes have exactly the same intrinsic topological order, $SU(3)_1\cong SO(3)_2$. They remain distinct phases through their realization of the global $U(1)_B$ symmetry. In the infrared, condensing $\phi_{(\mathbf{6},2)}$ gives the SC$^*$ phase, i.e. superconductivity response with a decoupled $SO(3)_2$, 
\begin{equation}
    \CL_{SO(3)_2}+ \frac{1}{2\pi}c d( \tr\widetilde a-B)+\frac{2}{2\pi}b d \tr\widetilde a \rightarrow \CL_{SO(3)_2}+\frac{2}{2\pi}b dB
\end{equation}
where $b$ is the dynamical $U(1)$ gauge field. The gauge-invariant superconducting order parameter $\Delta=\det\phi_R$ has physical charge two and can be interpreted as a bound state of three charge-$2e/3$ anyons, giving a strong-pairing interpretation \cite{read2000strongpairing,fan2026anyonsc,Shi:2026oxz,Wang:2026ziw,han2026anyonsc,Mehta:2026vnh}.
Thus $\langle\Delta\rangle\neq0$ breaks $U(1)_B$ to $\mathbb Z_2$. The remaining $\IZ_2$ global symmetry acts trivially on $SO(3)_2$, which is consistent with the trivial fractionalization class $H^2(\IZ_2,\IZ_3)=1$. Our construction provides a Chern-Simons-Higgs description of the transition into the anyon-superconducting phase, complementing previous non-Abelian parton constructions~\cite{Wang:2026ziw,Mehta:2026vnh}.

This construction extends to $SU(N)_1$ for $N\geq3$ due to the conformal embedding $SO(N)_2\hookrightarrow SU(N)_1$ \cite{Schellekens:1986mb,Bais:1986zs,Aharony:2016jvv,Lim:2024qis,Zhang:2025xkw}. Using its $U(N)$ presentation, we add a scalar $\phi_R$ in the representation $R=(\mathrm{Sym}^2(\mathbf N),2)$, with $\tr\widetilde a=B$. A potential favoring $\phi_R=v\mathbf1_N$ gives the phase diagram
\begin{equation}
\begin{tikzpicture}[
baseline=(current bounding box.center),
  x=1pt,y=1pt,
  every node/.style={
    inner sep=0pt,
    outer sep=0pt,
    anchor=base
  },
  phase/.style={font=\fontsize{10}{12}\selectfont},
  note/.style={font=\fontsize{9}{11}\selectfont},
  branch/.style={
    -{Stealth[length=3.4pt,width=2.5pt]},
    line width=0.4pt
  }
]
  \node[phase] at (0,0) {$SU(N)_1+\phi_{\mathbf R}$};
  \node[note] at (0,-12)
    {${\mathbf R}=(\mathrm{Sym}^2(\mathbf N),2)$};

  \draw[branch] (-17,-19) -- (-53,-34);
  \draw[branch] (17,-19) -- (53,-34);
  \node[note] at (-59,-25) {$m^2<0$};
  \node[note] at (59,-25) {$m^2>0$};

  \node[phase] at (-53,-45) {$SO(N)_2$};
  \node[phase] at (0,-45) {$\cong$};
  \node[phase] at (53,-45) {$SU(N)_1$};

  \node[note] at (-53,-59)
    {$U(1)_B\to\mathbb Z_2$ ($SC^*$)};
  \node[note] at (53,-59)
    {$U(1)_B$ preserved};
\end{tikzpicture}
\end{equation}
The two regimes share the intrinsic topological order $SO(N)_2\cong SU(N)_1$, with $N$ Abelian anyons obeying $\IZ_N$ fusion~\cite{Aharony:2016jvv}. The scalar has physical charge $2/N$, while $\det\phi_R$ has charge two. Its condensation spontaneously breaks $U(1)_B$ to $\IZ_2$, coexisting with the same intrinsic topological order \footnote{For odd $N$, no electric center one-form symmetry survives, whereas for even $N$ the scalar preserves $\IZ_2^{(1)}$, generated by $-\mathbf1_N$. This element lies in the residual $SO(N)$ gauge group and does not give an additional decoupled $\IZ_2$ gauge sector (in contrast with the $\Spin(8)_1$ example discussed in the End Matter). This one-form symmetry is distinct from the unbroken zero-form subgroup $\IZ_2\subset U(1)_B$.}.

For odd $N$, the un-Higgsed phase admits a fermionic $1/N$ Laughlin interpretation, with the transparent fermion and the invertible sectors required by level-rank duality summarized in \SMref{SM:LevelRankDuality} \cite{Hsin:2016blu}. For even $N$, a bosonic realization can instead describe a chiral spin liquid, with $U(1)_{B}$ interpreted as a spin-rotation symmetry. The same construction in the Higgs phase then describes spin-symmetry breaking coexisting with chiral topological order.

\textit{\textbf{Non-Integrable Representations.} ---} A curious feature of conformal embeddings is that none of the representations containing a singlet under the subgroup is also integrable at level 1 (see~\SMref{SM:LevelRankDuality}). It is therefore not immediately clear what interpretation, if any, the transition admits in terms of the anyons of the un-Higgsed phase.

Wilson lines in non-integrable representations are of course still meaningful in the UV theory. Moreover, it is reasonable to assume that the scalar in such a representation becomes light at the transition point and that its worldlines are energetically favored, driving the transition. This is partly why in the earlier example we were able to diagnose a transition even when the two topological orders agreed: although the un-Higgsed phase realizes the $\mathbb{Z}_{3}$ center symmetry of $SU(3)_{1}$, this is broken by the Wilson lines in the $\mathbf{6}$ that proliferate at the transition point. Because the topological order is broken there, the response to $U(1)_{B}$ is allowed to differ between the two phases, which is exactly the result obtained in the previous subsection.

More generally, we may guess the braiding properties of these non-topological Wilson lines by comparing them with their image after tuning the parameters to the un-Higgsed phase, which we explain in more detail below. In turn, this suggests which one-form symmetries are broken and which remain topological across the transition. This remains a guess, however, since there could be additional light modes at the strongly coupled transition point that could break lines that would otherwise remain topological.

\textit{\textbf{Screening Argument.} ---}
Adjoint screening provides a candidate identification of non-integrable scalar representations with IR anyon sectors in the un-Higgsed phase. It offers a complementary perspective on the transition between the topological sectors $SU(3)_1$ and $SO(3)_2$.

For $m^2\gg0$, the infrared topological order is $SU(3)_1$. The scalar in $\mathbf6$ admits the screening channel $ \mathbf6 \otimes\mathbf8\supset\overline{\mathbf3}$. Dressing by adjoint gauge fluctuations preserves both its center charge and its physical electric charge $2/3$. Since $\overline{\mathbf3}$ is the unique level-one anyon with this center charge, it is a natural candidate for the infrared sector of the dressed scalar excitation.

The candidate anyon generates the entire category in the un-Higgsed phase, $\mathcal{B} = \langle\overline{\mathbf{3}}\rangle = SU(3)_{1}$, and its M\"uger centralizer is trivial, $\mathcal{B}'_{SU(3)_{1}} = \mathrm{Vec}$. Since no non-trivial line braids trivially with $\overline{\mathbf{3}}$, the $\IZ_{3}$ one-form symmetry is broken during the transition~\cite{Cheng:2026hgu}, consistent with the complete breaking of the electric center one-form symmetry by triality-two matter.

For $m^2\ll0$, the Higgs phase is $SO(3)_2$. Under the unbroken gauge group, the scalar representation branches as $\mathbf6=\mathbf1 \oplus \mathbf5$. The $\mathbf5$ labels an integrable Abelian boson in the covering theory $\Spin(3)_2\cong SU(2)_4$. Gauging the $\mathbb Z_2$ one-form symmetry generated by this boson identifies it with the vacuum in $SO(3)_2=SU(2)_4/\mathbb Z_2$, while the anyon $\mathbf3$ splits into two nontrivial Abelian lines. Thus, the $\mathbf5$ becomes trivial in the Higgs phase and will not constitute an additional anyon condensation.

This argument extends to the $SU(N)_1\rightarrow SO(N)_2$ family for $N\geq3$. The un-Higgsed phase is straightforward from the center charge. For the Higgs phase, the scalar representation branches in the unbroken gauge group $SO(N)$ as $\mathrm{Sym}^2(\mathbf N)=\mathbf1\oplus S$, where $S$ is the symmetric traceless tensor representation. The $S$ labels an integrable Abelian boson in $\Spin(N)_2$ and generates the $\IZ_2$ one-form symmetry gauged in passing to $SO(N)_2=\Spin(N)_2/\IZ_2$. Thus, $S$ is identified with the vacuum in the Higgs phase.

More generally, adjoint screening of a scalar in representation $R$ gives channels $R\otimes A=\bigoplus_i R_i$, where $A=\mathrm{Adj}(G)$. Since $A$ is center-neutral, all channels retain the center charge of $R$, including those reached by repeated screening in $R\otimes A^{\otimes m}$.

An energetic heuristic for selecting a screening channel comes from a short-distance color-exchange interaction $H_{\mathrm{ex}}=\lambda T_R^aT_A^a$ with $\lambda>0$. In a channel $T\subset R \otimes A$, its contribution is $E_{\mathrm{ex}}(T)
    =\frac{\lambda}{2}
    \bigl[C_2(T)-C_2(R)-C_2(A)\bigr]$. 
For single adjoint screening, this interaction favors the channel with the smallest quadratic Casimir. Assuming that exchange dominates the channel ordering and that the binding gain compensates the dressing cost, we propose the heuristic choice 
\begin{equation} \label{CasimirCriterion}
    R^\circ\in
    \underset{T\subset R \otimes A}{\operatorname{arg\,min}}\,C_2(T).
\end{equation}
For $SU(N)_1$ with $R=\mathrm{Sym}^2(\mathbf N)$, the smallest Casimir channel is $\wedge^2\mathbf N\subset R \otimes A$, which is also the unique level-one anyon with center charge two.

If the matter field has trivial center charge, then it preserves the electric center one-form symmetry. For example, a $\mathbf{27}$ scalar in $SU(3)_1$ can screen to the vacuum through $\mathbf{27}\rightsquigarrow\mathbf8\rightsquigarrow\mathbf1$, and its proliferation does not break the one-form symmetry. Indeed, an $SO(3)$-invariant VEV has full stabilizer $SO(3)\times\IZ_3$, yielding $SO(3)_2\boxtimes\CD(\IZ_3)$, where $\CD(\IZ_3)$ is the untwisted $\IZ_3$ gauge theory. The original anomalous $\mathbb Z_3^{(1)}$ symmetry remains exact and can be represented by the dyonic subgroup generated by $em$ within $\CD(\mathbb Z_3)$.

Matching the center charge is of course a necessary but not sufficient condition for identifying the infrared anyon after screening. For gauge groups with trivial center, this condition imposes no constraint, and the identification requires additional dynamical input. The smallest-Casimir criterion proposed above remains a heuristic.

\textit{\textbf{Non-Abelian Anyon Proliferation.} ---}
As noted above, conformal embeddings also yield Higgsing transitions in which the un-Higgsed phase can be obtained from the Higgsed one by non-Abelian anyon condensation. The condensable algebra need not contain every fusion channel of its constituents, as established in earlier studies of Hopf symmetry breaking, anyon condensation, and gapped interfaces~\cite{BaisSchroersSlingerland2003,Kong:2013aya,LanWangWen2015}. Here we present an explicit example connecting such Higgsing transitions to anyon proliferation in a twisted $S_{3}$ quantum double, consistent with the minimal condensation principle of Reference \cite{Lu:2026fid}. Further examples can be constructed from other conformal embeddings. A list of all conformal embeddings can be found in \cite{davydov2013witt}.

Consider $\Spin(16)_1$ coupled to a scalar $\phi$ in the spinorial representation $\mathbf{13312}_s$ \footnote{A smaller irrep $\mathbf{5304}$ has an additional unbroken $\IZ_2$ center. The unbroken gauge group in this case is $\Spin(9)\times \IZ_2$.}. A potential favoring a $\Spin(9)$-invariant condensate gives the full stabilizer $\Spin(9)$ and the Higgs flow
\begin{equation} \label{Spin9Higgsing}
    \Spin(16)_1+\phi_{\mathbf{13312}_s}
    \xrightarrow{m^2<0}\Spin(9)_2.
\end{equation}
Both sides have the same chiral central charge $c_-=8$. At the level of intrinsic topological order, $\Spin(16)_1$ is equivalent to the toric code $\CD(\IZ_2)$, while $\Spin(9)_2$ is equivalent to a twisted quantum double $\CD(S_3)^\omega$ with $\omega \cong[4]\in H^3(S_3,U(1))\cong \IZ_6$ \cite{bischoff2018cat,ardonne2016classification}. The conformal embedding relates them by (recall \eqref{ConformalEmbeddingDualityGlobalForm})
\begin{equation} \label{Spin9condensation}
    \Spin(9)_2/\CA\cong\Spin(16)_1,
    \qquad \CA=\mathbf1\oplus \rC,
\end{equation}
where $\rC$ is the non-Abelian electric charge associated with the two-dimensional irrep of $S_3$. Alternatively, it is the Wilson line transforming in the $\mathbf{84}$ of $\Spin(9)$. However:
\begin{equation}
    \rC\otimes \rC=\mathbf1 \oplus \rB\oplus \rC,
\end{equation}
with $\rB$ the Abelian charge associated with the sign representation, corresponding to the Wilson line transforming in the $\mathbf{44}$ of $\Spin(9)$. Although $\rB$ is a gaugable boson appearing in $\rC \otimes \rC$, it is absent from $\CA$ and survives the condensation as an order-two boson in $\Spin(16)_{1}$. Thus, the proliferation transition admits a minimal condensation channel containing $\rC$ without condensing the full Lagrangian algebra generated by $\rC$.

The condensation \eqref{Spin9condensation} suggests that the Higgsing \eqref{Spin9Higgsing} describes a proliferation transition in which the $\rC$ anyon (the $\mathbf{84}$) of $\Spin(9)$ becomes light and proliferates. Note that this is not inconsistent with the fact that the gauge group of the CS-matter theory is $\Spin(16)$, since the latter is only a choice of description for our microscopic theory. Other dual descriptions (see e.g. \cite{Hsin:2016blu, Aharony:2016jvv, Cordova:2017vab, Cordova:2018qvg}) could make manifest other degrees of freedom, such as the anyons of $\Spin(9)$. In such a description, it would presumably be easier to diagnose the proliferating anyon directly from the Lagrangian.

To see how the proliferation suggested by \eqref{Spin9condensation} manifests itself in the $\Spin(16) + \phi_{\mathbf{13312}_{s}}$ CS-matter theory, notice that under the unbroken gauge group the non-integrable scalar representation branches as
\begin{equation} \label{13312branching}
\mathbf{13312}_{s} = \mathbf{1} \oplus \mathbf{36} \oplus \mathbf{44} \oplus2 (\mathbf{84}) \oplus \cdots.
\end{equation}
That is, the anyon $\rC$ (the $\mathbf{84}$) of $\Spin(9)$ appears in the branching rule of the Higgs field. Thus, if we interpret \eqref{Spin9condensation} as the anyon $\rC$ becoming light and proliferating, the branching rule implies that the Higgs field contains a corresponding light mode, precisely reproducing the fact that the scalar becomes light at the transition point \footnote{This fact is common to other conformal embeddings, suggesting indeed that condensing anyons drive the Higgs to become light at the transition.}.

Finally, let us analyze what happens to the Wilson line in the $\mathbf{13312}_{s}$ as we tune the parameters to the un-Higgsed phase instead. Adjoint gauge fluctuations allow screening $\mathbf{13312}_s\rightsquigarrow\mathbf{128}_s$. The scalar is neutral under a $\IZ_2$ subgroup of the $\Spin(16)$ center, which coincides with the center of the residual $\Spin(9)$ gauge group. The corresponding $\IZ_2^{(1)}$ symmetry remains exact, and its generator $B$ is deconfined in both phases. The non-Abelian sectors in the Higgs phase emerge from the residual gauge dynamics.

\textit{\textbf{Discussion and Outlook.} ---} 
In this work, we studied the relation between non-Abelian Higgsing transitions and anyon proliferation using symmetry enrichment, anyon condensation, and screening arguments. Symmetry enrichment distinguishes phases with equivalent intrinsic topological orders, while screening relates non-integrable scalar representations to candidate infrared anyon sectors in the un-Higgsed phase. The branching rule of the Higgs representation ties the two descriptions together: the anyons that become light in the condensation picture are carried by modes of the Higgs field, which becomes light precisely as $m^{2}$ is tuned from positive to negative values. Within this proliferation picture, the M\"uger centralizer of the proliferating subcategory describes the one-form symmetry inherited from the topological order in the un-Higgsed phase, while the Higgs phase generically contains additional topological lines emerging from the residual gauge dynamics.

Such emergence is not restricted to non-integrable scalar representations. For example, an integrable scalar in the $\mathbf6$ of $SU(3)_2$ can Higgs the theory to $SO(3)_4$ \cite{Cordova:2025eim}. At the level of braided categories, $SU(3)_2\cong\overline{\mathrm{Fib}}\boxtimes SU(3)_{-1}$ and $SO(3)_4\cong\overline{\mathrm{Fib}}\boxtimes\overline{\mathrm{Fib}}$, where $\overline{\mathrm{Fib}}$ denotes the Fibonacci category with reversed braiding. The scalar is directly identified with anyon $\mathbf6$, which generates the $SU(3)_{-1}$ factor, whose M\"uger centralizer is the $\overline{\mathrm{Fib}}$ sector generated by $\mathbf8$. Thus, one $\overline{\mathrm{Fib}}$ factor survives the transition, while the second emerges in the Higgs phase.

It would also be interesting to study transitions associated with holomorphic conformal embeddings, which likewise lead to phases related by anyon condensation~\cite{Cordova:2025zkz}.

\begin{acknowledgments}
\textit{\textbf{Acknowledgments.}} We would like to thank Meng Cheng, Ho Tat Lam, Ashvin Vishwanath, and Taige Wang for useful discussions. DGS is supported by the Society of Fellows at Harvard University. DGS also acknowledges support by the US Department of Energy under grant DE-SC0007870 and the Simons Collaboration on Celestial Holography. D.C.L. is supported by the Simons Collaboration on Ultra-Quantum Matter, which is a grant from the Simons Foundation (grant No. 651440, A.V.). We acknowledge use of ChatGPT 5.6 and 6.0, and Claude for making tables and figures, editorial assistance, and proofreading.
\end{acknowledgments}

\bibliographystyle{apsrev4-1}
\bibliography{main}

\newpage

\section*{End Matter}\label{sec:endmatter}

In this End Matter we provide a few more examples of Higgsing transitions based on conformal embeddings. These include the ACE $PSU(3) \subset \Spin(8)$, illustrating the importance of keeping track of disconnected sectors of the gauge group, as well as two examples showcasing non-Abelian anyon condensation and its relation to the branching rules of the scalar representation. More examples are presented in the Supplemental Material.

\textit{\textbf{Spin(8) ACE Example.} ---}
The bosonic $\mathrm{Spin}(8)_1$ theory realizes the three-fermion topological order, its nontrivial anyons $\psi_{v,s,c}$ are Abelian fermions with $\mathbb Z_2\times\mathbb Z_2$ fusion and pairwise mutual braiding phase $-1$~\cite{Burnell:2014ThreeFermion, Chen:2017Orbifolding}. Its chiral central charge $c_-=4$ requires time-reversal breaking in a strictly two-dimensional bosonic realization. An $S_3$ triality symmetry permutes the three fermions; gauging it promotes the non-Abelian twist defects to dynamical anyons~\cite{Barkeshli:2014cna,Teo:2015TwistLiquids,
Cui:2016Gauging,Chen:2017Orbifolding,Lu:2024lzf}.

We consider adding a real scalar in $\mathbf{56}_v=\Lambda^3\mathbf8_v$, which permits a Higgsing transition. We choose its potential to favor a $PSU(3)$-invariant condensate with full stabilizer $PSU(3)\times\mathbb Z_2^v$, where $\mathbb Z_2^v$ is the center subgroup acting trivially on $\mathbf{56}_v$~\cite{Witt:2008Triality}. It Higgses to,
\begin{equation}
\begin{aligned}
    \mathcal T_{\mathrm H}
    &=PSU(3)_3\boxtimes\mathcal D(\mathbb Z_2)\cong\mathrm{Spin}(8)_1\boxtimes\mathcal D(\mathbb Z_2),
\end{aligned}
\label{eq:spin8_56v_higgs}
\end{equation}
Here $PSU(3)_3=SU(3)_3/\mathbb Z_3^{(1)}$, and the equivalence follows from the conformal embedding $A_{2,3}\subset D_{4,1}$~\cite{Yu:2021zmu}. The toric-code factor $\mathcal D(\mathbb Z_2)$ arises from the residual $\mathbb Z_2^v$ gauge subgroup. The corresponding Higgs flow is
\begin{equation}
\begin{tikzpicture}[
  x=1pt,y=1pt,
  every node/.style={
    inner sep=0pt,
    outer sep=0pt,
    anchor=base
  },
  phase/.style={font=\fontsize{10}{12}\selectfont},
  note/.style={font=\fontsize{9}{11}\selectfont},
  branch/.style={
    -{Stealth[length=3.4pt,width=2.5pt]},
    line width=0.4pt
  }
]
  \node[phase] at (0,0)
    {$\mathrm{Spin}(8)_1+\phi_{\mathbf R}$};
  \node[note] at (0,-12)
    {${\mathbf R}=\mathbf{56}_v=\Lambda^3\mathbf8_v$};

  \draw[branch] (-17,-19) -- (-58,-34);
  \draw[branch] (17,-19) -- (58,-34);
  \node[note] at (-62,-25) {$m^2<0$};
  \node[note] at (62,-25) {$m^2>0$};

  \node[phase] at (-58,-45)
    {$PSU(3)_3\boxtimes\mathcal D(\mathbb Z_2)$};
  \node[phase] at (58,-45)
    {$\mathrm{Spin}(8)_1$};

  \node[note] at (-58,-59)
    {$\cong\mathrm{Spin}(8)_1
      \boxtimes\mathcal D(\mathbb Z_2)$};
\end{tikzpicture}
\label{eq:spin8}
\end{equation}
We consider screening of the non-integrable $\mathbf{56}_v$ by adjoint fluctuations $\mathbf{56}_v \times\mathbf{28}\supset\mathbf8_v$. 
The minimum-Casimir channel $\mathbf8_v$ preserves the scalar's center charge and corresponds to $\psi_v$ under our screening prescription. Its fusion subcategory $\CB=\langle \psi_v\rangle= \mathrm{sVec}$ and M\"uger centralizer obey $\mathcal B'_{\mathrm{Spin}(8)_1} \simeq\mathrm{sVec}$ which suggests a surviving fermionic $\mathbb Z_2$ one-form symmetry~\cite{Cheng:2026hgu}. The toric-code sector provides a minimal modular extension of $\mathrm{sVec}$.

To introduce $U(1)_B$ enrichment, we use
\begin{equation}
    \mathrm{Spin}^c(8)
    =\frac{\mathrm{Spin}(8)\times U(1)}
    {\langle(z_s,-1)\rangle},
\end{equation}
where $z_s$ generates $\mathbb Z_2^s$ \footnote{One can similarly use the quotient generated by $(z_c,-1)$.}. Writing the added $U(1)$ connection as $a_0$, the physical connection is $\widetilde a=2a_0$. The coupling $\frac{1}{2\pi}c\,d(\widetilde a-B)$ fixes $\widetilde a=B$, so a field with charge $q$ under the added $U(1)$ has physical charge $q/2$.
 
We introduce a complex scalar $\Phi$ in $(\mathbf{56}_v,1)$, with physical charge $1/2$, and choose its potential to favor the same $PSU(3)$-invariant condensate. The quadratic gauge singlet $\Delta=\frac{1}{3!}\Phi_{abc}\Phi^{abc}$ has physical charge 1. Then it acquires an expectation value and breaks $U(1)_B$ completely. For this bosonic topological order, the $U(1)_B$ can be thought of as the spin-rotation symmetry. In particular, the infrared Lagrangian is given by
\begin{equation}
\begin{aligned}
    \CL_{PSU(3)_3}+\frac{2}{2\pi}a_v d b+\frac{1}{2\pi}u d B+ \cdots
\end{aligned}
\end{equation}
Here $a_v$ and $b$ are compact $U(1)$ gauge fields describing the residual $\mathbb Z_2^v$ gauge sector, while $u$ is particle-vortex dual to the Higgs field.
This phase therefore supports a Goldstone mode together with $PSU(3)_3\boxtimes\mathcal D(\mathbb Z_2)$ topological order.

Triality acts on the gauge field by outer automorphisms and permutes $\mathbf{56}_{v,s,c}$. We therefore consider an $S_3$-symmetric theory with three real scalars, 
\begin{equation}
\begin{aligned}
    &\mathrm{Spin}(8)_1
    +\phi_{\mathbf{56}_v}
    +\phi_{\mathbf{56}_s}
    +\phi_{\mathbf{56}_c}\\
    &\qquad\longrightarrow
    PSU(3)_3\cong\mathrm{Spin}(8)_1.
\end{aligned}
\end{equation}
An $S_3$-invariant potential can favor condensates aligned along singlet directions of a common $PSU(3)$ subgroup. Their stabilizers $PSU(3)\times\mathbb Z_2^i$ intersect in $PSU(3)$, eliminating the additional toric-code sector.

\textit{\textbf{Example: $A_{1,28} \hookrightarrow G_{2,1}$} ---} We now move to additional examples involving conformal embeddings that are not Abelian, for which the two phases are related by non-Abelian anyon condensation. Higgsing transitions of this kind have recently been considered, e.g. in \cite{Schafer-Nameki:2026hgx, Cheng:2026hgu}, as a way of constructing continuum Lagrangians describing the proliferation of non-Abelian anyons.

We first study an example where the ambient CS interaction is $G_{2,1}$. The un-Higgsed phase thus corresponds to the Fibonacci topological order. We pick the embedding $SO(3) \subset G_{2}$. Recalling that $SO(3)_{k} \equiv SU(2)_{2k}/\mathbb{Z}_{2}$~\cite{Aharony:2016jvv}, we see that the CS gauge group of the Higgs phase is $SO(3)_{14}$. The smallest $G_{2}$ representation containing an $SO(3)$ singlet is the $\mathbf{77}'$ \footnote{The fundamental has Dynkin labels $(1,0$). The $\mathbf{77}$ has Dynkin labels $(3,0)$, while the $\mathbf{77}'$ has $(0,2)$.}:
\begin{equation} \label{77'branchingrule}
    \mathbf{77}' = \mathbf{1} \oplus \mathbf{5} \oplus \mathbf{9} \oplus \mathbf{11} \oplus \mathbf{13} \oplus \mathbf{17} \oplus \mathbf{21}.
\end{equation}
Coupling $G_{2,1}$ to a scalar in the $\mathbf{77}'$ along with an interaction potential such that the VEV stabilizes the $SO(3)$, we find the RG flow diagram:
\begin{equation}
\begin{tikzpicture}[
  baseline=(current bounding box.center),
  x=1pt,y=1pt,
  every node/.style={
    inner sep=0pt,
    outer sep=0pt,
    anchor=base
  },
  phase/.style={font=\fontsize{10}{12}\selectfont},
  note/.style={font=\fontsize{9}{11}\selectfont},
  branch/.style={
    -{Stealth[length=3.4pt,width=2.5pt]},
    line width=0.4pt
  }
]
  \node[phase] at (0,0) {$G_{2,1}+\phi_{\mathbf{77}'}$};

  \draw[branch] (-14,-8) -- (-48,-27);
  \draw[branch] (14,-8) -- (48,-27);
  \node[note] at (-53,-15) {$m^2<0$};
  \node[note] at (53,-15) {$m^2>0$};

  \node[phase] at (-48,-36) {Higgs};
  \node[phase] at (-48,-47) {$SO(3)_{14}$};
  \node[phase] at (48,-36) {Un-Higgsed};
  \node[phase] at (48,-47) {$G_{2,1}$};
\end{tikzpicture}
\label{higgsing_flow_G2_SO3}
\end{equation}
Since the embedding is conformal, we can relate the two topological orders by non-Abelian anyon condensation (recall Eqn. \eqref{ConformalEmbeddingDualityGlobalForm}):
\begin{equation} \label{nacondensation}
    \frac{SO(3)_{14}}{\mathcal{A}} \cong G_{2,1}.
\end{equation}
Specifically, the condensable algebra is given by
\begin{equation} \label{CondensableAlgebra}
    \mathcal{A} = \mathbf{1} \oplus \mathbf{11},
\end{equation}
where we have denoted the anyons according to the dimension of the $SO(3)$ representation \footnote{Notice that the lines in the spin-5 and spin-9 representations of $SU(2)_{28}$ are identified after gauging the center to obtain $SO(3)_{14}$.}.

The condensable algebra \eqref{CondensableAlgebra} suggests that the Higgsing \eqref{higgsing_flow_G2_SO3} describes a proliferation transition in which the anyon $\mathbf{11}$ of $SO(3)_{14}$ becomes light, proliferates, and drives the system into the Fibonacci topological order $G_{2,1}$ (recall the paragraph above Eqn. \eqref{13312branching} in the main text). As in the $\Spin(9) \subset \Spin(16)$ example, we can see an avatar of this interpretation by analyzing the branching rule of the non-integrable scalar representation \eqref{77'branchingrule}, in which the $\mathbf{11}$ clearly appears. Therefore, if we interpret \eqref{CondensableAlgebra} as the $\mathbf{11}$ becoming light and proliferating, the branching rule implies that the scalar field has a corresponding light mode, reproducing the fact that the scalar becomes light at the transition point.

We end this example by analyzing the fate of the Wilson line in the $\mathbf{77}'$ as we drive the theory into the un-Higgsed phase. The $\mathbf{77}'$ is non-integrable in $G_{2,1}$, so we must consider its screening. Taking the tensor product of the $\mathbf{77}'$ with the adjoint $\mathbf{14}$, we find
\begin{equation}
    \mathbf{77}' \otimes \mathbf{14} = \mathbf{14} \oplus \cdots,
\end{equation}
so the representation with the smallest Casimir is the adjoint itself. The tensor product of $\mathbf{14}$ with itself gives
\begin{equation}
    \mathbf{14} \otimes \mathbf{14} = \mathbf{1} \oplus \cdots,
\end{equation}
and the smallest Casimir is now carried by the trivial representation $\mathbf{1}$. Thus, we expect the light Wilson lines in the $\mathbf{77}'$ at the transition point to be transparent with respect to the Fibonacci line in the un-Higgsed phase.

\textit{\textbf{Example: $A_{1,10} \hookrightarrow B_{2,1}$} ---} We finish this End Matter with another example of non-Abelian anyon proliferation, based on the group embedding $SU(2) \subset \Spin(5)$. The un-Higgsed phase corresponds to $\Spin(5)_{1}$, while the Higgs phase is described by $SU(2)_{10}$.

The smallest $\Spin(5)$ representation containing an $SU(2)$ singlet is the $\mathbf{30}$ (the rank-three symmetric representation), which is non-integrable at level 1. The branching rule for this representation is
\begin{equation} \label{HiggsBranching}
    \mathbf{30} = \mathbf{1} \oplus \mathbf{7} \oplus \mathbf{9} \oplus \mathbf{13}.
\end{equation}
Thus, coupling $\Spin(5)_{1}$ to a scalar in the $\mathbf{30}$ and turning on a potential such that the VEV is stabilized by the $SU(2)$ subgroup, we obtain the following RG flow:
\begin{equation}
\begin{tikzpicture}[
  baseline=(current bounding box.center),
  x=1pt,y=1pt,
  every node/.style={
    inner sep=0pt,
    outer sep=0pt,
    anchor=base
  },
  phase/.style={font=\fontsize{10}{12}\selectfont},
  note/.style={font=\fontsize{9}{11}\selectfont},
  branch/.style={
    -{Stealth[length=3.4pt,width=2.5pt]},
    line width=0.4pt
  }
]
  \node[phase] at (0,0) {$\Spin(5)_1+\phi_{\mathbf{30}}$};

  \draw[branch] (-14,-8) -- (-48,-27);
  \draw[branch] (14,-8) -- (48,-27);
  \node[note] at (-53,-15) {$m^2<0$};
  \node[note] at (53,-15) {$m^2>0$};

  \node[phase] at (-48,-37) {Higgs};
  \node[phase] at (-48,-47) {$SU(2)_{10}$};
  \node[phase] at (48,-37) {Un-Higgsed};
  \node[phase] at (48,-47) {$\Spin(5)_1$};
\end{tikzpicture}
\label{higgsing_flow_Spin5}
\end{equation}

The topological order in the un-Higgsed phase can be obtained from that of the Higgsed phase by anyon condensation, which, as in the previous examples, is guaranteed by the conformality of the embedding (recall Eqn. \eqref{ConformalEmbeddingDualityGlobalForm}). More precisely, 
\begin{equation} 
    \frac{SU(2)_{10}}{\mathcal{A}} \cong \Spin(5)_{1},
\end{equation}
where the condensable algebra is
\begin{equation} \label{nacondensationSU2}
    \mathcal{A} = \mathbf{1} \oplus \mathbf{7},
\end{equation}
and we have denoted the anyons according to the dimension of the $SU(2)$ representation.

The condensable algebra \eqref{nacondensationSU2} suggests that the phase diagram \eqref{higgsing_flow_Spin5} describes a proliferation transition in which the anyon $\mathbf{7}$ of $SU(2)_{10}$ becomes light, proliferates, and drives the system into $\Spin(5)_{1}$ (recall the paragraph above Eqn. \eqref{13312branching} in the main text). A manifestation of this interpretation is visible in the branching rule of the scalar representation, Eqn. \eqref{HiggsBranching}. Indeed, the $\mathbf{7}$ appears in this branching. Thus, if we interpret the anyon condensation \eqref{nacondensationSU2} as the anyon $\mathbf{7}$ becoming light and proliferating, the branching implies that the scalar field becomes light through the corresponding mode, as expected at the transition point.

Finally, we analyze the Wilson line in the $\mathbf{30}$ as we move to the un-Higgsed phase. Since the $\mathbf{30}$ is non-integrable, we must study its screening. As we will see, in this example the Casimir criterion (Eqn. \eqref{CasimirCriterion} of the main text) plays a non-trivial role in identifying the candidate anyon of the un-Higgsed phase to which the $\mathbf{30}$ screens. Taking the tensor product with the adjoint $\mathbf{10}$, we find
\begin{equation}
    \mathbf{30} \otimes \mathbf{10} = \mathbf{30} \oplus \mathbf{35} \oplus \mathbf{81} \oplus \mathbf{154}.
\end{equation}
Interestingly, although the $\mathbf{30}$ has the smallest dimension in the tensor product, the representation with the smallest Casimir is actually the $\mathbf{35}$:
\begin{equation}
    C_{2}(\mathbf{30}) = 18, \quad \text{while} \quad C_{2}(\mathbf{35}) = 12.
\end{equation}
The representation energetically favored by the screening is thus the $\mathbf{35}$. Screening the latter further gives
\begin{equation}
    \mathbf{35} \otimes \mathbf{10} = \mathbf{5} \oplus \mathbf{10} \oplus \mathbf{14} \oplus \mathbf{30} \oplus 2(\mathbf{35}) \oplus \mathbf{35}' \oplus \mathbf{81} \oplus \mathbf{105},
\end{equation}
and the energetically preferred representation is now the $\mathbf{5}$. No further screening occurs, since
\begin{equation}
    \mathbf{5} \otimes \mathbf{10} = \mathbf{5} \oplus \mathbf{10} \oplus \mathbf{35},
\end{equation}
and thus the candidate anyon describing the $\mathbf{30}$ in the un-Higgsed phase is the $\mathbf{5}$.

\clearpage
\onecolumngrid

\setcounter{section}{0}
\setcounter{equation}{0}
\setcounter{figure}{0}
\setcounter{table}{0}

\renewcommand{\thesection}{S\arabic{section}}
\renewcommand{\theequation}{S\arabic{equation}}
\renewcommand{\thefigure}{S\arabic{figure}}
\renewcommand{\thetable}{S\arabic{table}}

\makeatletter

\begin{center}
\textbf{\large Supplemental Material}
\end{center}

\setcounter{table}{0}
\renewcommand{\thetable}{S\arabic{table}}
\setcounter{figure}{0}
\renewcommand{\thefigure}{S\arabic{figure}}
\setcounter{equation}{0}
\renewcommand{\theequation}{S\arabic{equation}}

\section{I. Abelian Conformal Embeddings For Exceptional Ambient Group}\label{SM:AbelianConformalEmbeddings} 

In this section, we discuss Higgsing transitions based on ACEs with an exceptional ambient gauge group. Many interesting topological orders are realized in this manner. For instance, the CS theories $G_{2,1}$ and $F_{4,1}$ correspond to the Fibonacci topological order and its time reversal, respectively. Likewise, $E_{6,1}$ and $E_{7,1}$ describe the time reversals of the $SU(3)_{1}$ and $SU(2)_{1}$ (also called semion) topological orders, respectively~\cite{Cordova:2018qvg}. Finally, $E_{8,1}$ corresponds to a bosonic invertible phase: the Kitaev $E_{8}$ state. The following CS-matter theories therefore provide examples in which such topological orders appear on both sides of the Higgsing.

Notice that conformal embeddings have been fully classified \cite{Bais:1986zs, Schellekens:1986mb}, so it is straightforward to find embeddings into ambient exceptional groups that are Abelian. For completeness, a list summarizing this class of conformal embeddings is presented in Table \ref{TableofExceptionalEmbeddings}. \\

\textit{\textbf{Example: $A_{1,1} \oplus A_{1,3} \hookrightarrow G_{2,1}$} ---} We begin with an example involving the Fibonacci topological order described by $G_{2,1}$. The Higgsing transition corresponds to the unique ACE into $G_{2,1}$: $SO(4) \subset G_{2}$. Keeping track of the levels, the appropriate global form of the connected gauge group after Higgsing is:
\begin{equation} \label{IRFibonacci}
     \frac{SU(2)_1 \boxtimes SU(2)_3}{\mathbb{Z}_{2}},
\end{equation}
where the quotient is the common center in between the two $SU(2)$ factors. Because of the duality implied by the conformal embeddings pointed out in the main text, the resulting topological order is equivalent to the Fibonacci topological order $G_{2,1}$. (This is straightforward to check in practice performing the quotient above explicitly following the three-step gauging rule of \cite{Moore:1989yh, Hsin:2018vcg}.)

To perform the Higgsing transition, we couple the CS term $G_{2,1}$ to a scalar field transforming in the $\mathbf{27}$ (the rank two symmetric-traceless representation), which is the smallest representation allowing a singlet under the $SO(4)$ subgroup:
\begin{equation}
    \mathbf{27} = (\mathbf{1}, \mathbf{1}) \oplus (\mathbf{2}, \mathbf{4}) \oplus (\mathbf{2}, \mathbf{2}) \oplus (\mathbf{3}, \mathbf{3})  \oplus (\mathbf{1}, \mathbf{5}).
\end{equation}
We assume an interaction potential exists that Higgses the gauge group to its connected component without additional discrete factors. See Appendix A of \cite{Cordova:2018qvg} for an argument of how such potentials are constructed.

 The resulting RG flow diagram takes the form:
\begin{equation}
\begin{tikzpicture}[
  baseline=(current bounding box.center),
  x=1pt,y=1pt,
  every node/.style={
    inner sep=0pt,
    outer sep=0pt,
    anchor=base
  },
  phase/.style={font=\fontsize{10}{12}\selectfont},
  note/.style={font=\fontsize{9}{11}\selectfont},
  branch/.style={
    -{Stealth[length=3.4pt,width=2.5pt]},
    line width=0.4pt
  }
]
  \node[phase] at (0,0) {$G_{2,1}+\phi_{\mathbf{27}}$};

  \draw[branch] (-14,-8) -- (-60,-27);
  \draw[branch] (14,-8) -- (60,-27);
  \node[note] at (-67,-15) {$m^2<0$};
  \node[note] at (67,-15) {$m^2>0$};

  \node[phase] at (-60,-37) {Higgs};
  \node[phase] at (-60,-47)
    {$\bigl(SU(2)_1\boxtimes SU(2)_3\bigr)/\mathbb Z_2
      \cong G_{2,1}$};
  \node[phase] at (60,-37) {Un-Higgsed};
  \node[phase] at (60,-47) {$G_{2,1}$};
\end{tikzpicture}
\label{G2SO4Embedding}
\end{equation}

We now discuss what happens to the Wilson line in the $\mathbf{27}$ as we tune the parameters out of the transition point. Since $G_{2}$ has trivial center, we cannot use $n$-ality to identify the candidate anyon line in the un-Higgsed phase. The screening interpretation still provides a non-trivial candidate anyon, which can be exposed by taking iterated tensor products of the adjoint $\mathbf{14}$ with the $\mathbf{27}$. The representation with the smallest Casimir in the tensor product is the fundamental $\mathbf{7}$:
\begin{equation}
    \mathbf{27} \otimes \mathbf{14} = \mathbf{7} \oplus \cdots.
\end{equation}
However, a second iteration taking the tensor product of the adjoint with the fundamental gives:
\begin{equation}
    \mathbf{7} \otimes \mathbf{14} = \mathbf{7} \oplus \cdots
\end{equation}
and thus the $\mathbf{7}$ is again the representation with the smallest Casimir. It is then clear that further tensoring with the adjoint will always return the $\mathbf{7}$. The screening interpretation therefore suggests that the Wilson line at the transition point transforming in the $\mathbf{27}$ becomes the Fibonacci anyon as we move out of the critical point. In particular, we expect the topological order to be broken across the Higgsing, i.e. the two Fibonacci topological orders are not continuously connected as we vary the parameters of the UV theory.

\begin{table*}[t]
\centering
\renewcommand{\arraystretch}{1.8}
\begin{tabular}{|p{3.5cm}|p{3.5cm}|p{2cm}|}
\hline 
\multicolumn{3}{|c|}{Abelian Conformal Embeddings for Exceptional Lie Groups} \\
\hline
Affine Embedding & Global Form & $\mathbf{R}_{\mathrm{min}}$ \\
\hline
$A_{1,3}\oplus A_{1,1} \hookrightarrow G_{2,1}$ & $\frac{SU(2)_{3} \boxtimes SU(2)_{1}}{\mathbb{Z}_{2}}$ & $\mathbf{27}$ \\
$A_{1,1}\oplus C_{3,1} \hookrightarrow F_{4,1}$ & $\frac{SU(2)_{1} \boxtimes USp(6)_{1}}{\mathbb{Z}_{2}}$ & $\mathbf{324}$ \\
$A_{2,2}\oplus A_{2,1} \hookrightarrow F_{4,1}$ & $\frac{SU(3)_{2} \boxtimes SU(3)_{1}}{\mathbb{Z}_{3}}$ & $\mathbf{273}$ \\
$A_{1,1}\oplus A_{5,1} \hookrightarrow E_{6,1}$ & $\frac{SU(2)_{1} \boxtimes SU(6)_{1}}{\mathbb{Z}_{2}}$ & $\mathbf{650}$ \\
$A_{2,1}^{\,3} \hookrightarrow E_{6,1}$ & $\frac{SU(3)_{1} \boxtimes SU(3)_{1} \boxtimes SU(3)_{1}}{\mathbb{Z}_{3}}$ & $\mathbf{650}$ \\
$C_{4,1} \hookrightarrow E_{6,1}$ & $PUSp(8)_{1}$ & $\mathbf{351}'$ \\
$A_{1,1}\oplus D_{6,1} \hookrightarrow E_{7,1}$ & $\frac{SU(2)_{1} \boxtimes \Spin(12)_{1}}{\mathbb{Z}_{2}}$ & $\mathbf{1539}$ \\
$A_{2,1}\oplus A_{5,1} \hookrightarrow E_{7,1}$ & $\frac{SU(3)_{1} \boxtimes SU(6)_{1}}{\mathbb{Z}_{3}}$ & $\mathbf{1463}$ \\
$A_{7,1} \hookrightarrow E_{7,1}$ & $\frac{SU(8)_{1}}{\mathbb{Z}_{2}}$ & $\mathbf{1463}$ \\
$A_{1,1}\oplus E_{7,1} \hookrightarrow E_{8,1}$ & $\frac{SU(2)_{1} \boxtimes E_{7,1}}{\mathbb{Z}_{2}}$ & $\mathbf{3875}$  \\
$A_{2,1}\oplus E_{6,1} \hookrightarrow E_{8,1}$ & $\frac{SU(3)_{1} \boxtimes E_{6,1}}{\mathbb{Z}_{3}}$ & $\mathbf{30380}$ \\
$A_{4,1}\oplus A_{4,1} \hookrightarrow E_{8,1}$ & $\frac{SU(5)_{1} \boxtimes SU(5)_{1}}{\mathbb{Z}_{5}}$ & $\mathbf{30380}$ \\
$D_{8,1} \hookrightarrow E_{8,1}$ & $\frac{\Spin(16)_{1}}{\mathbb{Z}_{2}}$ & $\mathbf{27000}$ \\
$A_{8,1} \hookrightarrow E_{8,1}$ & $\frac{SU(9)_{1}}{\mathbb{Z}_{3}}$ & $\mathbf{30380}$ \\
\hline
\end{tabular}
\caption{List of ACEs with exceptional ambient group. The embedding of affine Lie algebras is presented along with the global form of the subgroup. $\mathbf{R}_{\mathrm{min}}$ in the third column stands for the smallest irreducible representation of the ambient group with a singlet, and such that a generic stabilizer contains only the connected subgroup without additional discrete contributions.}  \label{TableofExceptionalEmbeddings}
\end{table*}

\textit{\textbf{Example: $A_{2,1} \oplus A_{2,2} \hookrightarrow F_{4,1}$} ---} The $F_{4,1}$ CS theory is the time reversal of the Fibonacci topological order ($F_{4,1} \cong G_{2,-1}$)~\cite{Cordova:2018qvg}. The group embedded in $F_{4}$ has the global form $\left(SU(3) \times SU(3)\right)/\mathbb{Z}_{3}$. Including the CS levels, the IR Higgs gauge group is given by
\begin{equation}
    \frac{SU(3)_{2} \boxtimes SU(3)_{1}}{\mathbb{Z}_{3}},
\end{equation}
where the quotient is by the $\mathbb{Z}_{3}$ center common to the two $SU(3)$ factors. Following the main text, we deduce that the resulting topological order is equivalent to that of the un-Higgsed CS theory $F_{4,1}$.

The smallest representation of $F_{4}$ which has a singlet under the $(SU(3) \times SU(3)) /\mathbb{Z}_{3}$ subgroup is the $\mathbf{273}$:
\begin{equation}
    \mathbf{273} = (\mathbf{1}, \mathbf{1}) \oplus (\mathbf{3}, \mathbf{3}) \oplus (\overline{\mathbf{3}}, \overline{\mathbf{3}}) \oplus \cdots
\end{equation}
Thus, we couple the $F_{4,1}$ CS term to a scalar in the $\mathbf{273}$ and add a generic interaction potential. We find the RG flow diagram:
\begin{equation}
\begin{tikzpicture}[
  baseline=(current bounding box.center),
  x=1pt,y=1pt,
  every node/.style={
    inner sep=0pt,
    outer sep=0pt,
    anchor=base
  },
  phase/.style={font=\fontsize{10}{12}\selectfont},
  note/.style={font=\fontsize{9}{11}\selectfont},
  branch/.style={
    -{Stealth[length=3.4pt,width=2.5pt]},
    line width=0.4pt
  }
]
  \node[phase] at (0,0) {$F_{4,1}+\phi_{\mathbf{273}}$};

  \draw[branch] (-14,-8) -- (-60,-27);
  \draw[branch] (14,-8) -- (60,-27);
  \node[note] at (-67,-15) {$m^2<0$};
  \node[note] at (67,-15) {$m^2>0$};

  \node[phase] at (-60,-37) {Higgs};
  \node[phase] at (-60,-47)
    {$\bigl(SU(3)_1\boxtimes SU(3)_2\bigr)/\mathbb Z_3
      \cong F_{4,1}$};
  \node[phase] at (60,-37) {Un-Higgsed};
  \node[phase] at (60,-47) {$F_{4,1}$};
\end{tikzpicture}
\label{F4SU3SU3Embedding}
\end{equation}

We now discuss what happens to the Wilson line in the $\mathbf{273}$ as we flow to the un-Higgsed phase. As in the previous example, since $F_{4,1}$ has trivial center, we cannot use $n$-ality to identify the candidate anyon. We may still identify a candidate studying the screening. Performing iterated tensor products of the adjoint $\mathbf{52}$ with the $\mathbf{273}$, we find the representation with the smallest Casimir in the tensor product is the fundamental $\mathbf{26}$:
\begin{equation}
    \mathbf{273} \otimes \mathbf{52} = \mathbf{26} \oplus \cdots.
\end{equation}
Taking the tensor product of the adjoint $\mathbf{52}$ with the $\mathbf{26}$ gives:
\begin{equation}
    \mathbf{26} \otimes \mathbf{52} = \mathbf{26} \oplus \cdots
\end{equation}
and thus $\mathbf{26}$ is again the representation with the smallest Casimir. The screening interpretation suggests therefore that as we tune the parameters to the un-Higgsed phase, the $\mathbf{26}$ becomes the Fibonacci anyon of $F_{4,1}$. In particular, we expect the topological order to be broken throughout the Higgsing, i.e. the two Fibonacci topological orders are not continuously connected as we vary the UV parameters.

\textit{\textbf{Example: $A_{1,1} \oplus C_{3,1} \hookrightarrow F_{4,1}$} ---} We now discuss yet another example involving the time reversal of Fibonacci, $F_{4,1}$. We will see, however, that the screening argument now gives a different answer than in the previous example: the Wilson line in the representation of the Higgs field is completely screened. The global form of the group embedded in $F_{4}$ is $(SU(2) \times USp(6))/ \mathbb{Z}_{2}$. Keeping track of the levels, the global form of the CS gauge group in the Higgs phase is
\begin{equation}
    \frac{SU(2)_1 \boxtimes USp(6)_1}{\mathbb{Z}_{2}},
\end{equation}
where the quotient is the common center in between the $SU(2)$ and the $USp(6)$ factor. The resulting topological order is equivalent to $F_{4,1}$.

The $\mathbf{324}$ is the smallest representation allowing a  $(SU(2) \times USp(6))/ \mathbb{Z}_{2}$ singlet:
\begin{equation}
    \mathbf{324} = (\mathbf{1}, \mathbf{1}) \oplus (\mathbf{1}, \mathbf{14}) \oplus (\mathbf{2}, \mathbf{14}') \oplus (\mathbf{3}, \mathbf{21}) \oplus (\mathbf{2}, \mathbf{64}) \oplus (\mathbf{1}, \mathbf{90}).
\end{equation}
Coupling $F_{4,1}$ to a scalar in this representation and adding a generic Higgsing potential, we find the RG flow diagram:
\begin{equation}
\begin{tikzpicture}[
  baseline=(current bounding box.center),
  x=1pt,y=1pt,
  every node/.style={
    inner sep=0pt,
    outer sep=0pt,
    anchor=base
  },
  phase/.style={font=\fontsize{10}{12}\selectfont},
  note/.style={font=\fontsize{9}{11}\selectfont},
  branch/.style={
    -{Stealth[length=3.4pt,width=2.5pt]},
    line width=0.4pt
  }
]
  \node[phase] at (0,0) {$F_{4,1}+\phi_{\mathbf{324}}$};

  \draw[branch] (-14,-8) -- (-60,-27);
  \draw[branch] (14,-8) -- (60,-27);
  \node[note] at (-67,-15) {$m^2<0$};
  \node[note] at (67,-15) {$m^2>0$};

  \node[phase] at (-60,-37) {Higgs};
  \node[phase] at (-60,-47)
    {$\bigl(SU(2)_1\boxtimes USp(6)_1\bigr)/\mathbb Z_2
      \cong F_{4,1}$};
  \node[phase] at (60,-37) {Un-Higgsed};
  \node[phase] at (60,-47) {$F_{4,1}$};
\end{tikzpicture}
\label{F4SU2USp6Embedding}
\end{equation}

To see that the Wilson line in the $\mathbf{324}$ is completely screened, we follow the main text. We first take the tensor product of the adjoint $\mathbf{52}$ with the $\mathbf{324}$. The representation with the smallest Casimir in the tensor product is the adjoint $\mathbf{52}$:
\begin{equation}
    \mathbf{324} \otimes \mathbf{52} = \mathbf{52} \oplus \cdots.
\end{equation}
Next, we iterate and take the tensor product of the adjoint $\mathbf{52}$ with itself, which contains the trivial representation:
\begin{equation}
    \mathbf{52} \otimes \mathbf{52} = \mathbf{1} \oplus \cdots.
\end{equation}
The $\mathbf{324}$ is therefore completely screened, and the topological order plausibly remains unbroken as we vary the parameters of the UV Lagrangian.

\textit{\textbf{Example: $A_{8,1} \hookrightarrow E_{8,1}$} ---} The $E_{8,1}$ CS theory is a bosonic invertible phase: the Kitaev $E_{8}$ state. To exemplify Higgsings in which the topological orders of the two semiclassical regimes are trivial and described by this invertible phase, we study the group embedding $SU(9)/\mathbb{Z}_{3} \hookrightarrow E_{8}$. The Higgs phase is:
\begin{equation}
    \frac{SU(9)_1}{\mathbb{Z}_{3}},
\end{equation}
and the duality implied by the conformal embedding guarantees that $SU(9)_1/\mathbb{Z}_{3} \cong E_{8,1}$ as topological orders, which is also easy to verify by taking the quotient explicitly. Notice that the MTC data of $SU(9)_{1}$ is equivalent to that of twisted $\mathbb{Z}_{3}$ gauge theory (see Section 4 of \cite{Cordova:2025zkz}). The fact that $SU(9)_1/\mathbb{Z}_{3} \cong E_{8,1}$ is the statement that condensing a $\mathbb{Z}_{3}$ generator completely trivializes the MTC data. (This observation generalizes to $SU(n^{2})_{1}$ for any odd integer $n$.)

The smallest representation of $E_{8}$ which has a singlet under the $SU(9)/\mathbb{Z}_{3}$ subgroup, and such that no contributions to the Higgsed gauge group exist besides the connected component is the $\mathbf{30380}$ \footnote{Notice that the $\mathbf{27000}$ also allows an $SU(9)/\mathbb{Z}_{3}$ singlet, but it is even under charge conjugation. Charge conjugation is an element of the ambient group $E_{8}$ that does not lie in the $SU(9)/\mathbb{Z}_{3}$ subgroup, so the IR gauge group is enhanced by a discrete factor. This is easily remedied by taking the $\mathbf{30380}$ instead, which allows a singlet that is odd under charge conjugation. The IR is then fully described by the connected part $SU(9)/\mathbb{Z}_{3}$, consistent with the assumption made throughout the text.}:
\begin{equation}
    \mathbf{30380} = \mathbf{1} \oplus \mathbf{84} \oplus \overline{\mathbf{84}} \oplus \cdots.
\end{equation}
Thus, we couple the $E_{8}$ CS term to a scalar in the $\mathbf{30380}$ and add a generic interaction potential that Higgses $E_{8}$ down to $SU(9)/\mathbb{Z}_{3}$. We find the RG flow diagram:
\begin{equation}
\begin{tikzpicture}[
  baseline=(current bounding box.center),
  x=1pt,y=1pt,
  every node/.style={
    inner sep=0pt,
    outer sep=0pt,
    anchor=base
  },
  phase/.style={font=\fontsize{10}{12}\selectfont},
  note/.style={font=\fontsize{9}{11}\selectfont},
  branch/.style={
    -{Stealth[length=3.4pt,width=2.5pt]},
    line width=0.4pt
  }
]
  \node[phase] at (0,0) {$E_{8,1}+\phi_{\mathbf{30380}}$};

  \draw[branch] (-14,-8) -- (-60,-27);
  \draw[branch] (14,-8) -- (60,-27);
  \node[note] at (-67,-15) {$m^2<0$};
  \node[note] at (67,-15) {$m^2>0$};

  \node[phase] at (-60,-37) {Higgs};
  \node[phase] at (-60,-47)
    {$SU(9)_1/\mathbb Z_3\cong E_{8,1}$};
  \node[phase] at (60,-37) {Un-Higgsed};
  \node[phase] at (60,-47) {$E_{8,1}$};
\end{tikzpicture}
\label{E8SU9Embedding}
\end{equation}

We conclude that the Higgs phase and the un-Higgsed phase both have trivial topological order. It is easy to check that the screening argument identifies the $\mathbf{30380}$ with the identity line.

\textit{\textbf{Example: $D_{8,1} \hookrightarrow E_{8,1}$} ---} We finish with another example involving the $E_{8,1}$ invertible phase. We take the embedding of the group $\Spin(16)/\mathbb{Z}_{2}^{s}$ into $E_{8}$. Keeping track of the levels, we find that the Higgsed CS theory is $\frac{\Spin(16)_{1}}{\mathbb{Z}^{s}_{2}}$. Note that this quotient is a different global form from $SO(16)_{1}$, which is always a fermionic TQFT. The quotient corresponds to the $\mathbb{Z}_{2}$ subgroup of the center under which the vectorial and cospinorial representations are non-trivially charged, while the spinorial representation is neutral. Recalling that the MTC data of $\Spin(16)_{1}$ is the same as that of the $\mathbb{Z}_{2}$ toric code, it is straightforward to check that the anyon content of $\Spin(16)_{1}/\mathbb{Z}^{s}_{2}$ is trivial. Thus, as topological orders
\begin{equation}
    \frac{\Spin(16)_{1}}{\mathbb{Z}^{s}_{2}} \cong E_{8,1},
\end{equation}
which is a rather simple instance of the dualities generically implied by \eqref{AbelianDuality}.

The smallest representation of $E_{8}$ which has a singlet under the $\Spin(16)/\mathbb{Z}_{2}^{s}$ subgroup is the $\mathbf{27000}$:
\begin{equation}
    \mathbf{27000} = \mathbf{1} \oplus \mathbf{128} \oplus \mathbf{1820} \oplus \mathbf{5304} \oplus \overline{\mathbf{6435}} \oplus \mathbf{13312}.
\end{equation}
Thus, we couple the $E_{8}$ CS term to a scalar in the $\mathbf{27000}$ and add a generic interaction potential. We find the RG flow diagram:
\begin{equation}
\begin{tikzpicture}[
  baseline=(current bounding box.center),
  x=1pt,y=1pt,
  every node/.style={
    inner sep=0pt,
    outer sep=0pt,
    anchor=base
  },
  phase/.style={font=\fontsize{10}{12}\selectfont},
  note/.style={font=\fontsize{9}{11}\selectfont},
  branch/.style={
    -{Stealth[length=3.4pt,width=2.5pt]},
    line width=0.4pt
  }
]
  \node[phase] at (0,0) {$E_{8,1}+\phi_{\mathbf{27000}}$};

  \draw[branch] (-14,-8) -- (-60,-27);
  \draw[branch] (14,-8) -- (60,-27);
  \node[note] at (-67,-15) {$m^2<0$};
  \node[note] at (67,-15) {$m^2>0$};

  \node[phase] at (-60,-37) {Higgs};
  \node[phase] at (-60,-47)
    {$\Spin(16)_1/\mathbb Z_2^s\cong E_{8,1}$};
  \node[phase] at (60,-37) {Un-Higgsed};
  \node[phase] at (60,-47) {$E_{8,1}$};
\end{tikzpicture}
\label{higgsing_flow_E8_SO16}
\end{equation}
so the topological order is trivial on both phases and described by the $E_{8}$ Kitaev state.

\section{II. Duality between $SU(n)_1$ and $U(1)_{-n}$}
\label{SM:LevelRankDuality}

For odd $n$, the $U(1)_n$ theory is fermionic, requiring spin structure of the manifold, while $SU(n)_1$ is bosonic. Therefore, the relation between $SU(n)_1$ and the Abelian Laughlin theory requires specifying both the transparent fermion and the invertible topological sector. We review the level-one duality of Hsin and Seiberg~\cite{Hsin:2016blu} and express it in the conventions of the main text. In particular, $\widetilde a$ denotes a $U(n)$ connection and the compact $U(1)$ gauge field $c$ imposes $\tr\widetilde a=B$. We first work on a fixed spin three-manifold and take $B$ to be an ordinary background $U(1)$ connection. Wedge products are implicit, and the gauge trace is taken in the fundamental representation of $SU(n)$.

The hat in $\widehat{SU(n)_1}$ denotes the spin theory obtained by adding a transparent fermion $f$ to $SU(n)_1$. A Lagrangian for this additional sector is
\begin{equation}
    \mathcal L_0[x,y]
    =\frac{1}{4\pi}x\,dx+\frac{1}{2\pi}x\,dy = \frac{1}{4\pi} \begin{pmatrix}x&y \end{pmatrix}K_0  d\begin{pmatrix}x\\y \end{pmatrix},
    \qquad
    K_0=\begin{pmatrix}1&1\\1&0\end{pmatrix},
\label{eq:sm_transparent_fermion}
\end{equation}
where $x$ and $y$ are compact $U(1)$ gauge fields. The matrix $K_0$ is unimodular and has zero signature. Thus this sector has one ground state on every closed spatial surface and vanishing chiral central charge, while retaining the local fermion line. The hat therefore changes the line content but does not supply the gravitational counterterm needed for level-rank duality.

We use the gravitational normalization of Ref.~\cite{Hsin:2016blu},
\begin{equation}
    \mathrm{CS}_{\mathrm{grav}}
    =\frac{1}{192\pi}\tr\left(\omega\,d\omega+\frac{2}{3}\omega^3\right),
\label{eq:sm_gravitational_cs}
\end{equation}
where $\omega$ is the gravitational connection and the trace in this equation is over tangent-space indices. The displayed expression is a local representative of the globally defined spin Chern--Simons functional. In this convention, $-2\mathrm{CS}_{\mathrm{grav}}$ represents an invertible phase with $c_-=1$.

To retain the background convention of the main text, define
\begin{equation}
    \mathcal L_n[\widetilde a,c;B]
    =\frac{1}{4\pi}\tr\left(\widetilde a\,d\widetilde a
    -\frac{2i}{3}\widetilde a^3\right)
    +\frac{1}{2\pi}c\,d(\tr\widetilde a-B).
\label{eq:sm_unitary_constraint_action}
\end{equation}
The change of variables $u=\widetilde a+c\mathbf1_n$ gives, up to a total derivative,
\begin{equation}
\begin{aligned}
    \mathcal L_n[\widetilde a,c;B]
    ={}&\frac{1}{4\pi}\tr\left(u\,du-\frac{2i}{3}u^3\right)
    -\frac{n}{4\pi}c\,dc-\frac{1}{2\pi}c\,dB.
\end{aligned}
\label{eq:sm_level_rank_factorization}
\end{equation}
The first term describes the invertible spin theory $U(n)_1$. Its Abelian description and gravitational response are
\begin{equation}
    \frac{1}{4\pi}\tr\left(u\,du-\frac{2i}{3}u^3\right)
    \longleftrightarrow
    \sum_{j=1}^{n}\frac{1}{4\pi}r_j\,dr_j
    \longleftrightarrow
    -2n\,\mathrm{CS}_{\mathrm{grav}},
\label{eq:sm_invertible_unitary_sector}
\end{equation}
where the $r_j$ are compact $U(1)$ gauge fields. Each term in the sum is an explicitly written $U(1)_1$ Lagrangian. These equivalences retain the transparent fermion and use a common spin structure.

Writing $b\equiv-c$ in \eqref{eq:sm_level_rank_factorization}, we obtain the duality
\begin{equation}
\begin{aligned}
    \mathcal L_n[\widetilde a,c;B]
    +\frac{1}{4\pi}x\,dx+\frac{1}{2\pi}x\,dy
    \quad\longleftrightarrow\quad
    &-\frac{n}{4\pi}b\,db+\frac{1}{2\pi}b\,dB
    +\frac{1}{4\pi}x'\,dx'+\frac{1}{2\pi}x'\,dy'
    -2n\,\mathrm{CS}_{\mathrm{grav}}.
\end{aligned}
\label{eq:sm_sun_uone_duality}
\end{equation}
Here $x',y'$ are an independent pair of compact $U(1)$ fields on the Abelian side. Equation~\eqref{eq:sm_sun_uone_duality} is Eqn.~(3.8) of Ref.~\cite{Hsin:2016blu}, with its auxiliary $\mathrm{spin}_c$ background set to zero and its ordinary background replaced by $-B$. We have subtracted the common counterterm $B\,dB/(4\pi)$ from both sides. This choice gives exactly the constraint and background contact-term convention in Eqn.~\eqref{eq:sm_unitary_constraint_action}. For odd $n$, the Abelian theory already contains a transparent fermion, so its displayed $x',y'$ sector is redundant. The odd-level $U(n)$ formulation in Eqn.~\eqref{eq:sm_unitary_constraint_action} is itself a spin theory; its transparent fermion must still be retained even if we use additional $c$ to set its trace equal to the background gauge field $B$.

\textit{The $SU(3)_1$ example.}
Specializing to $n=3$, and assigning the relative gravitational counterterm to the non-Abelian side, gives the duality between $SU(3)_1$ and $U(1)_{-3}$,
\begin{equation}
\begin{aligned}
    \mathcal L_{\mathrm{Ab}}[b;B]
    &=-\frac{3}{4\pi}b\,db+\frac{1}{2\pi}b\,dB,
    \\
    \mathcal L_{\mathrm{non\text{-}Ab}}[\widetilde a,c,x,y;B]
    &=\frac{1}{4\pi}\tr\left(\widetilde a\,d\widetilde a
    -\frac{2i}{3}\widetilde a^3\right)
    +\frac{1}{2\pi}c\,d(\tr\widetilde a-B)
    \\
    &\quad+\frac{1}{4\pi}x\,dx+\frac{1}{2\pi}x\,dy
    +\frac{1}{32\pi}\tr\left(\omega\,d\omega+\frac{2}{3}\omega^3\right),
    \\
    \mathcal L_{\mathrm{Ab}}[B]
    &\longleftrightarrow\mathcal L_{\mathrm{non\text{-}Ab}}[B].
\end{aligned}
\label{eq:sm_su3_laughlin_explicit}
\end{equation}
where $b,c,x,y$ are dynamical compact $U(1)$ gauge fields, $\widetilde a$ is the $U(3)$ gauge field, $B$ is the background $U(1)$ gauge field that tracks the $U(1)$ global symmetry, and $\omega$ is the spin connection. The $x,y$ terms explicitly retain the transparent fermion; they do not introduce an additional deconfined non-trivial topological sector. Equivalently, the same non-Abelian dual admits an entirely gauge-field presentation,
\begin{equation}
\begin{aligned}
    \mathcal L_{\mathrm{non\text{-}Ab}}'
    ={}&\frac{1}{4\pi}\tr\left(\widetilde a\,d\widetilde a
    -\frac{2i}{3}\widetilde a^3\right)
    +\frac{1}{2\pi}c\,d(\tr\widetilde a-B)
    \\
    &-\frac{1}{4\pi}v_1\,dv_1
    -\frac{1}{4\pi}v_2\,dv_2
    -\frac{1}{4\pi}v_3\,dv_3,
\end{aligned}
\label{eq:sm_su3_invertible_completion}
\end{equation}
where $v_1,v_2,v_3$ are compact $U(1)$ gauge fields. The three negative-level Abelian terms give the inverse of the invertible $U(3)_1$ sector in Eqn.~\eqref{eq:sm_level_rank_factorization} and contain a local fermion.

The line statistics and framing anomaly provide direct checks. Turning off the background gauge field, let $W_q=\exp(iq\oint b)$ denote the Abelian Wilson lines. Retaining the transparent fermion, $q$ is valued in $\mathbb Z_6$ and $W_3=f$. One choice of line map is
\begin{equation}
    \mathbf3\longleftrightarrow W_1f,
    \qquad
    \overline{\mathbf3}\longleftrightarrow W_2,
    \qquad
    f\longleftrightarrow W_3.
\label{eq:sm_su3_line_map}
\end{equation}
Indeed, $h_{W_q}=-q^2/6\pmod1$, so $h_{W_1f}=h_{W_2}=1/3$, matching the two nontrivial $SU(3)_1$ anyons. This is an intrinsic line map at $B=0$; fractional $U(1)_B$ charges match modulo integers, and matching specific charged representatives can require attachment of integer-charge local excitations. The invertible completion has $c_-=-3$, giving total chiral central charge $2-3=-1$, as required by $U(1)_{-3}$. Integrating out $b$ in Eqn.~\eqref{eq:sm_su3_laughlin_explicit} gives the local electromagnetic response
\begin{equation}
    \mathcal L_{\mathrm{response}}[B]
    =\frac{1}{12\pi}B\,dB,
\label{eq:sm_laughlin_response}
\end{equation}
and $W_q$ carries electric charge $q/3$. In particular, the screened $\overline{\mathbf3}$ channel of the $(\mathbf6,2)$ scalar corresponds to $W_2$, with charge $2/3$ and spin $1/3$.

Equation~\eqref{eq:sm_unitary_constraint_action} at $n=3$ is the topological gauge part of Eqn.~\eqref{eq:su3_symmetric_csh}. By itself it retains the invertible $U(3)_1$ contribution and has $c_-=2$. Identifying its un-Higgsed phase with the bare fermionic Laughlin theory therefore requires the fixed inverse sector displayed in Eqn.~\eqref{eq:sm_su3_invertible_completion}, or the equivalent gravitational counterterm in Eqn.~\eqref{eq:sm_su3_laughlin_explicit}. This sector is a spectator under the scalar deformation: it changes the absolute gravitational response but does not change the Higgs stabilizer or the charge-two order parameter $\det\phi_R$. The duality established here concerns the topological theories, and it is interesting to establish a duality of the critical theories.

We have worked at a fixed spin structure, with $B$ an ordinary $U(1)$ background. For a detailed treatment of $\mathrm{spin}_c$ connections and the electronic spin/charge relation, we refer to Ref.~\cite{Hsin:2016blu}.

\section{III. Conformal Embeddings have no Integrable Representations with a singlet}\label{SM:AbelianConformalEmbeddings}

Let $\mathfrak{h}_{\tilde{k}} \hookrightarrow \mathfrak{g}_{1}$ be a conformal embedding of affine Lie algebras. In this section of the Supplemental Material we establish an observation used throughout the main text: no integrable representation of $\mathfrak{g}_{1}$ can contain an $\mathfrak{h}_{\tilde{k}}$ singlet.

Assume, for contradiction, that such a singlet state $|s\rangle$ exists. This state furnishes a primary with respect to the smaller affine Lie algebra $\mathfrak{h}_{\tilde{k}}$. Moreover, as a singlet, $|s\rangle$ must be annihilated by the zero modes of the affine Lie subalgebra (i.e., the modes generating the ordinary Lie algebra $\mathfrak{h}$). We therefore have:\begin{equation}
    J^{a}_{(H), n} |s\rangle = 0, \quad n \geq 0,
\end{equation}
where $J^{a}_{(H), n}$ are the modes of the $\mathfrak{h}_{\tilde{k}}$ currents.

The $L^{\mathfrak{h}_{\tilde{k}}}_{0}$ Virasoro mode of the $\mathfrak{h}_{\tilde{k}}$ Sugawara stress tensor is
\begin{equation}
    L_0^{\mathfrak{h}_{\tilde{k}}} = \frac{1}{2(\tilde{k} + h_{\mathfrak{h}}^\vee)}\left( \sum_a J^{a}_{(H), 0} J^{a}_{(H),0} +
2\sum_{n=1}^{\infty}\sum_a J^{a}_{(H), -n}J^{a}_{(H), n} \right),
\end{equation}
where $h_{\mathfrak{h}}^{\vee}$ is the dual Coxeter number of $\mathfrak{h}$. Therefore $L_{0}^{\mathfrak{h}_{\tilde{k}}} |s\rangle = 0$ as well. Now, conformality of the embedding means that the Sugawara stress tensors of $\mathfrak{h}_{\tilde{k}}$ and $\mathfrak{g}_{1}$ coincide. In particular, $L_{0}^{\mathfrak{h}_{\tilde{k}}} = L_{0}^{\mathfrak{g}_{1}}$, and we deduce
\begin{equation}
    L_0^{\mathfrak{g}_{1}} |s\rangle = 0.
\end{equation}
However, the only state annihilated by the Virasoro zero mode of the ambient affine Lie algebra $\mathfrak{g}_{1}$ is the vacuum, yielding a contradiction. We therefore conclude that no integrable representation of $\mathfrak{g}_{1}$ can contain an $\mathfrak{h}_{\tilde{k}}$ singlet.

\end{document}